\documentclass[fleqn,usenatbib]{mnras}

\usepackage{newtxtext,newtxmath}

\usepackage[T1]{fontenc}

\DeclareRobustCommand{\VAN}[3]{#2}
\let\VANthebibliography\thebibliography
\def\thebibliography{\DeclareRobustCommand{\VAN}[3]{##3}\VANthebibliography}

\usepackage{graphicx}	
\usepackage{amsmath}	

\newcommand{\HI}{H{\sc i}}

\title[The distribution of the CNM in galaxy discs]{On the distribution of the Cold Neutral Medium in galaxy discs}

\author[Rowan J. Smith et al.]{
Rowan J.\ Smith,$^{1}$\thanks{E-mail: rowan.smith@manchester.ac.uk}
Robin Tress,$^{2}$
Juan D.\ Soler,$^{3}$
Ralf S.\ Klessen,$^{4,5}$
Simon C.\ O.\ Glover,$^{4}$
\newauthor
Patrick Hennebelle,$^{6}$
Sergio Molinari,$^{3}$
Mordecai-Mark Mac Low,$^{7}$
David Whitworth$^{4}$
\\ \\
$^{1}$Jodrell Bank Centre for Astrophysics, Department of Physics and Astronomy, University of Manchester, Oxford Road, Manchester M13 9PL, UK\\
$^{2}$\'Ecole Polytechnique F\'ed\'erale de Lausanne, Observatoire de Sauverny, Chemin Pegasi 51, 1290 Versoix, Switzerland\\
$^{3}$Istituto di Astrofisica e Planetologia Spaziali (IAPS). INAF. Via Fosso del Cavaliere 100, 00133 Roma, Italy \\
$^{4}$Universit\"{a}t Heidelberg, Zentrum f\"{u}r Astronomie, Institut f\"{u}r Theoretische Astrophysik, Albert-Ueberle-Stra{\ss}e 2, 69120 Heidelberg, Germany\\
$^{5}$Universit\"{a}t Heidelberg, Interdisziplin\"{a}res Zentrum f\"{u}r Wissenschaftliches Rechnen, Im Neuenheimer Feld 205, 69120 Heidelberg, Germany\\ 
$^{6}$ Laboratoire AIM, Paris-Saclay, CEA/IRFU/SAp - CNRS - Universit\'{e} Paris Diderot. 91191, Gif-sur-Yvette Cedex, France\\
$^{7}$ Department of Astrophysics, American Museum of Natural History, 200 Central Park West, New York, NY 10024, USA 
}

\date{Accepted XXX. Received YYY; in original form ZZZ}

\pubyear{2023}

\begin{document}
\label{firstpage}
\pagerange{\pageref{firstpage}--\pageref{lastpage}}
\maketitle

\begin{abstract}
The Cold Neutral Medium (CNM) is an important part of the galactic gas cycle and a precondition for the formation of molecular and star forming gas, yet its distribution is still not fully understood. In this work we present extremely high resolution simulations of spiral galaxies with time-dependent chemistry such that we can track the formation of the CNM, its distribution within the galaxy, and its correlation with star formation. We find no strong radial dependence between the CNM fraction and total \HI\ due to the decreasing interstellar radiation field counterbalancing the decreasing gas column density at larger galactic radii. However, the CNM fraction does increase in spiral arms where the CNM distribution is clumpy, rather than continuous, overlapping more closely with H$_2$. The CNM doesn’t extend out radially as far as \HI, and the vertical scale height is smaller in the outer galaxy compared to \HI\ with no flaring. The CNM column density scales with total midplane pressure and disappears from the gas phase below values of $P_T/k_B =1000$\,K\,cm$^{-3}$. We find that the star formation rate density follows a similar scaling law with CNM column density to the total gas Kennicutt-Schmidt law. In the outer galaxy we produce realistic vertical velocity dispersions in the \HI\ purely from galactic dynamics but our models do not predict CNM at the extremely large radii observed in \HI\ absorption studies of the Milky Way. We suggest that extended spiral arms might produce isolated clumps of CNM at these radii.

\end{abstract}

\begin{keywords}
ISM: structure -- ISM: kinematics and dynamics -- methods: numerical
\end{keywords}



\section{Introduction}

The interstellar medium (ISM) exists in a complex, multi-phase, form from hot ionised gas, to cold molecular star-forming clouds \citep[e.g.][]{McKee77,Draine11,Tielens2005, Klessen2016}. Matter cycles through these phases as stars formed in the cold dense molecular clouds release energetic feedback and momentum into their surroundings, which then influences the subsequent evolution of the galaxy by setting the equilibrium disc structure and depletion time.

A crucial component in this phase matter cycle is the cold neutral medium or CNM. This consists of the neutral atomic hydrogen (H{\sc i}) with temperatures around $100$\,K \citep{Kulkarni87,Dickey90} that makes up the bulk of the neutral gas in galaxies, alongside the warm neutral medium (WNM) which has temperatures of order $10^4$\,K. 
These atomic phases exist together in pressure equilibrium such that they can be considered as a two-phase medium \citep{Field69,Wolfire03,Bialy19}. It is from the CNM that gas is compressed and cooled to form individual molecular clouds where stars are born \citep[e.g.][]{Mckee2007, Girichidis20}. The CNM is consequently a gateway and a pre-condition for star formation in galaxies, and determining its distribution is crucial for any theory of galaxy evolution. In this paper, we seek to use numerical models to investigate the broad trends of where the CNM is located in galaxies, what sets this distribution, and how it corresponds to star formation.

The phases of the ISM are usually explained in terms of pressure equilibrium \citep{Field69,McKee77}. As summarised by \citet{Ostriker22}, the midplane pressure is in vertical dynamical equilibrium with the weight of the ISM. This midplane pressure then determines the balance between hot and two-phase (warm+cold) gas such that they have median pressures within 50\% of each other. Ultimately this leads to a near linear relationship with star formation ($\Sigma_{\rm{SFR}}$\,$\propto$\,$P^{1.2}$) due to the importance of feedback in setting this disc pressure. Similarly, the existence of the cold neutral gas phase should also be related to the local pressure environment. \citet{Wolfire03} have shown that there is a minimum thermal pressure for the CNM phase to coexist with the WNM, which has a value of $P_{\rm{th}}/k_B$\,$\approx$\,$3000$\,K\,cm$^{-3}$ at the solar circle. As the pressure falls with galactocentric radius, this leads to a predicted limit of $R$\,$<$\,$18$\,kpc for the CNM in the Milky Way.


One of the key ways of observing the CNM is through deep surveys of the absorption of the 21~cm line of neutral hydrogen against radio continuum sources since the emission is dominated by the warm phase. Using the Millenium Arecibo 21~cm absorption line survey, \citet{Heiles03} found that in the solar neighbourhood the CNM component makes up about $40\%$ of the neutral gas, whereas more recently \citet{Murray18} estimated the CNM was $28\%$ of the total H{\sc i} but that $20\%$ of the H{\sc i} was in a thermally unstable phase. \citet{Dickey09} combined multiple 21 cm emission surveys of the galactic plane to deduce that the CNM in the Milky Way was similarly located to the WNM. Moreover, recent work with the Australian Square Kilometer Array Pathfinder (ASKAP) by \citet{Dickey22} found that in the inner Milky Way, the CNM has a similar scale height to the molecular gas, but in the outer galaxy the CNM and WNM are well mixed and maintain a constant CNM/WNM ratio out to radii of at least 40 kpc. Similarly, \citet{Strasser07} found CNM at large galactocentric radii in spiral arms in the outer galaxy.

\citet{Soler22} investigated the filamentary structure in the H{\sc i} emission toward the Galactic disk using a Hessian matrix method. The identified filamentary structures correspond to roughly 80\% of the H{\sc i} emission and most likely consist of CNM material due to their higher density. The mean scale height of the filamentary H{\sc i}, was lower than that of the total H{\sc i} in the outer galaxy, suggesting that the CNM and WNM have different scale heights in this regime. These results are at odds with the aforementioned pressure equilibrium models, which suggest that in the low-pressure environment of the outer galaxy, the CNM phase should become increasingly less prevalent \citep{Wolfire03}.

Another way of probing the CNM phase is through the [C{\sc ii}] 158 $\mu$m emission line, as this is the main coolant of the diffuse ISM. As the emission is sensitive to the density of the gas, that from the WNM is a factor of $\sim 20$ less bright than that from the CNM. This means that [C{\sc ii}] 158\,$\mu$m emission from diffuse regions can be assigned to the CNM phase. This exercise was done by the GOTC++ \textit{Herschel}/HIFI survey published by \citet{PinedaJ13}. In contrast to H{\sc i} absorption studies these authors found that the CNM column density decreased more rapidly with galactocentric radius than that of the WNM, and consequently, the fraction of the atomic gas in the cold phase was much lower in the outer galaxy ($\sim 20\%$).

The distribution of the neutral atomic medium is something that has not just been a focus of observational studies, but is also a topic of interest for theoretical numerical studies. In cosmological simulations, resolving the scale height of the H{\sc i} gas is a key challenge for reproducing Milky Way type galaxies \citep[e.g][]{Hopkins18}. Previously, the H{\sc i} scale heights of simulated galaxies were of order of kiloparsecs \citep{Bahe16,Marinacci17}, however recent \textsc{FIREbox} \citep{Gensior22} simulations have more realistic heights of $\sim 100$\,pc in the galactic centres, rising to $\sim 800$\,pc in the outer galaxy. The authors postulate that the solution to faithfully reproducing galaxy scale heights comes from the inclusion of a realistic multiphase medium with cold gas and small scale stellar feedback. Such models typically omit a detailed chemical treatment of the gas, and even \textsc{FIREbox} has a mass resolution of over $6$\,$\times$\,$10^4$\,$M_\odot$. To obtain a finer description of the CNM, therefore, simulations of gas in an isolated galaxy are needed.

One approach is to simulate gas in stratified boxes \citep[e.g.][]{Hennebelle14b, Walch15, Girichidis16, Rathjen21}. However, these studies have mainly focused on molecular gas and star formation rather than H{\sc i}. One prominent example is the work of \citet{KimCG13} which showed that the star formation rate surface density varied almost linearly with the midplane pressure set by the weight of the ISM. As earlier stated, the local pressure balance of the ISM is also theorised to be extremely important in setting the CNM fraction \citep{Wolfire03}.

Unfortunately, stratified box simulations typically only cover an area of a few square kiloparsecs of an idealised disc and are therefore unable to investigate the full galactic distribution. One approach is to embed a co-rotating high-resolution box within a galaxy simulation as is done in the Cloud Factory simulations of \citet{Smith20}. These models, and stratified boxes from the FRIGG project \citep{Iffrig17}, have been used to interpret the orientations of the H{\sc i} filaments identified in The H{\sc i}/OH/Recombination-line survey of the inner Milky Way \citep[THOR,][]{Beuther16}, as reported in \citet{Soler20}. However, the simulations of \citet{Smith20} only reach a high resolution in a 3~kpc box in the star-forming disc. To investigate the true galactic distribution of the CNM we need the entire galaxy to be included. We, therefore, turn to an updated version of the isolated hydrodynamic galaxy models presented by \citet{Tress20a} and \citet{Tress21}, which reach parsec resolution or better in the cold gas across the galaxy while including full hydrogen chemistry.

The paper is structured as follows. We first outline the details of the galaxy models and how they are analysed in Section \ref{sec:methods}. Then we describe the radial and vertical distribution of the CNM that we find in Section \ref{sec:results}. In Section \ref{sec:discussion}, we investigate how this relates to the local pressure conditions in the galactic disc, and the effect of spiral arms. Finally, in Section \ref{sec:conclusions} we give our conclusions.


\section{Methods}\label{sec:methods}

\subsection{Isolated Galaxy Simulations}
\label{sec:sims} 
Our simulations are carried out using the \textsc{Arepo} code \citep{Springel10} with custom physics modules to describe star formation and cold dense gas. For a full description of our numerical setup, see \citet{Tress20a}. However, we briefly summarise the major features and differences from previous work here.

The models of \citet{Tress20a,Tress21} consist of two simulations of a galaxy disc consisting of dark matter halo ($6\times10^{11}$ M$_\odot$), bulge ($5.3\times10^{9}$ M$_\odot$), stellar disc ($4.77\times10^{10}$ M$_\odot$), and gas disc ($5.3\times10^{9}$ M$_\odot$). The first model is isolated and develops large scale spiral structure but with breaks and bifurcations, whereas the other is perturbed by a fly-by from a companion galaxy, inducing the formation of strong spiral arms. For our analysis, we focus on an updated version of the Isolated case as this best lends itself to radial averaging. However, in Section \ref{sec:spiral} we will revisit the latter Interacting model. 

\begin{figure}
 \includegraphics[width = 3.7in]{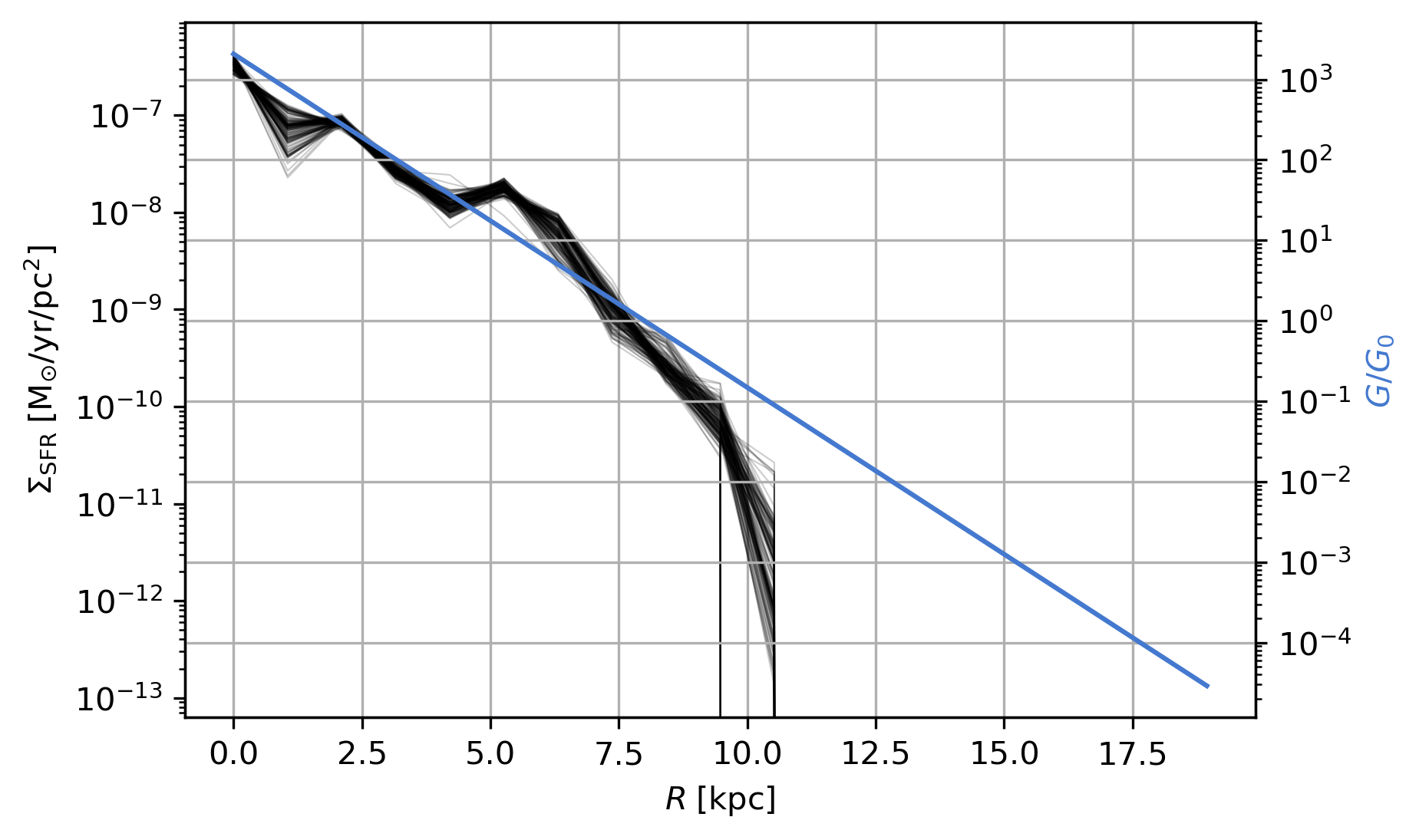}
 \caption{ The star formation rate density as a function of radius from the isolated galaxy simulation of \citet{Tress20a} between $280-320$ Myr. The blue line shows the exponential function used to represent the scaling of the interstellar radiation field. }
 \label{fig:uvfield}
\end{figure}

The original simulation assumed a constant interstellar radiation field consistent with the solar neighbourhood value $G_0$\,$=$\,$1.7$ in Habing units \citep{Draine78,Habing68}. However, this is not a good description of the outer galaxy where the field will be lower due to the low star formation rate. To account for this we computed the time-averaged radial star formation rate surface density profile from $280$ to $320$\,Myr in the original \citet{Tress20a} model during the steady state period of the galaxy. We then fit an exponential function to the star formation density and scale it such that, at the radius where the star formation rate surface density is equal to the solar neighbourhood value, it takes the value of $G_0$. The final expression takes the form 
\begin{equation}
G = G_0 \exp (-(R-R_0)/1.047), 
\end{equation}
where $G$ is the interstellar radiation field, $R$ the galactocentric radius in units of kpc, and $R_0$ the radius where the star formation surface density matches the value of the solar neighbourhood. Figure \ref{fig:uvfield} shows the average radial star formation rate density profiles from the original models, and the radial dependence of the new varying interstellar radiation field. (We checked subsequently that the star formation rate was not significantly different in the new runs with varying field.) Shielding from the field is computed assuming a constant shielding length of $30$\,pc using the \textsc{TreeCol} algorithm \citep{Clark12b}. We also assume that the mean free path of the FUV photons in the Milky Way is much less than the scale on which the SFR density is varying, this is a reasonable assumption out to galactocentric radii of 10 kpc in the Milky Way \citep{Wolfire03} but may be less valid in the outer galaxy. Nonetheless, while this is an approximation, it is far better than assuming a constant value for the interstellar radiation field. Cosmic ray ionisation is also included at a constant rate of $\zeta_H= 3\times10^{-17}$ s$^{-1}$ for atomic hydrogen, with the rates for other chemical species (H$_{2}$, He, etc.) scaled appropriately. However, in the atomic gas (i.e.\ when $A_V$\,$<$\,$1$), photoelectric heating dominates over cosmic ray heating by an order of magnitude or more (see e.g.\ Figure 10 in \citealt{Wolfire03}).

Heating and cooling of the gas is computed simultaneously with the solution of the chemical rate equations. We use the latest version of the cooling functions used by \citet{Clark19}.
A temperature floor of 20\,K is imposed on the ISM to prevent anomalously low temperatures occurring close to our resolution limit. 
For our chemistry, we adopt the approach first used in \textsc{Arepo} by \citet{Smith14a}, namely the NL97 network of \citet{Glover12a} which utilises the network for hydrogen chemistry first presented in \citet{Glover07a,Glover07b}. A simplified description of CO formation and destruction is also included in this network, based on \citet{Nelson97}. However, we do not analyse CO in this work.

For simplicity, we assume that the gas has solar metallicity throughout our simulated galaxy and that the dust-to-gas ratio is the same as in the solar neighbourhood. In reality, the Milky Way has a metallicity gradient of approximately $-0.037 \: {\rm dex \, kpc^{-1}}$ \citep{ArellanoCordova20}, and similar results are found in other nearby spiral galaxies \citep[see e.g.][]{Kreckel2019}. Therefore, in our simulation, the gas in the outer galaxy ($R > 12$~kpc) is roughly 2--3 times more metal-rich than we would expect to be the case in a real Milky Way-type spiral galaxy. However, this will have only a minor effect on the temperature of the atomic ISM, since the dominant cooling and heating processes have the same dependence on metallicity, provided that the dust-to-gas ratio scales linearly with the metallicity \citep{Wolfire95}.

The resolution of the simulations depends on two criteria. Firstly we set a target gas mass of 300\,$M_\odot$, which means that by default {\sc arepo} will refine or de-refine the grid so as to keep the masses of all of the grid cells within a factor of two of this value. However, on top of this we require that the Jeans length is resolved by at least four resolution elements to satisfy the Truelove criteria and avoid artificial fragmentation \citep{Truelove97, Greif11, Federrath11}. This leads to a mass resolution of $\sim 10\,$M$_\odot$ between densities of $10^{-22}$\,$<$\,$\rho$\,$<$\, $10^{-21}$\,g\,cm$^{-3}$, which equates to spatial scales of a parsec or smaller. We are therefore confident that the CNM is well resolved.

Star formation is modeled via sink particles \citep{Bate95,Federrath10}, which are non-gaseous particles that represent collapsing regions of gas that will form small (sub)clusters of stars. These are formed by checking if regions of gas with a density greater than $10^{-21}$\,g\,cm$^{-3}$ are unambiguously bound, collapsing, and accelerating inwards. \textit{Only} if these criteria are met will the gas be replaced with a sink particle, which can then accrete additional mass that falls within a radius of 2.5\,pc of the cell if it is gravitationally bound to it. As star formation is still inefficient at these scales we assume a 5\% star formation efficiency \citep{Evans09} and associate a stellar and gas fraction to each sink. 

Using the model of \citet{Sormani17}, we sample the IMF and associate supernovae with the massive stars as described by \citet{Tress20a}. For each supernova, we calculate an injection radius, which is the radius of the smallest sphere centred on the supernova that contains at least 40 grid cells. If the injection radius is smaller than the expected radius of a supernova remnant at the end of its Sedov-Taylor phase, we inject thermal energy from the supernova; otherwise, we inject momentum \citep[e.g.][]{Gatto15}. Mass is returned with each supernova explosion such that when the last supernova occurs the gaseous component of the sink is exhausted. The sink is then turned into a star particle. To account for type Ia supernovae, we also randomly select a star particle every 250 years and create a supernova event at its position 
\citep[based on the star formation history of M51, as quantified by][which is similar to our model]{Eufrasio17}.

\begin{figure}
 \includegraphics[width = 2.8in]{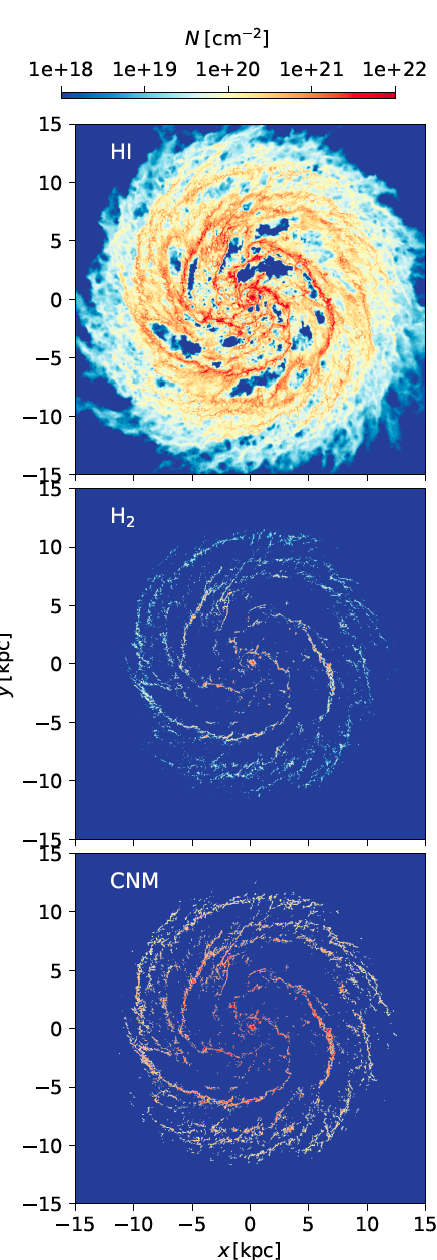}
 \caption{Projected column density of the isolated simulation in H{\sc i}, H$_2$, and the CNM, here defined to be any HI with a temperature below 200\,K. The CNM traces the distribution of the molecular gas rather than the full H{\sc i} distribution.}
 \label{fig:projection}
\end{figure}

Figure \ref{fig:projection} shows our updated Isolated galaxy simulation at a time of $300$ Myr, where we carry out our analysis. Note, that we performed a smooth averaging procedure on the original simulations to investigate if choosing a single snapshot in time biased the result. We found that the radial profile followed an identical trend radially but with a smoother shape. However, the averaging process masked real fluctuations due to spiral structure and so chose instead the snapshot analysis.


\subsection{Data Analysis}\label{sec:calculations}

In order to explore how the CNM varies as a function of galactocentric position and height above the midplane, we bin the simulation output in two different ways. Firstly, we bin the data in cylindrical co-ordinates in order to study whether the CNM fraction varies with galactocentric radius. We use 60 bins linearly from an inner radius of 0.1\,kpc to an outer one of 15\,kpc. In the Isolated galaxy simulation, the star formation is mostly contained within a radius $\sim 10$\,kpc and so both the inner and outer disc is included. To examine the impact of features such as spiral arms we also bin the data in angle using 60 bins from 0 to 360 degrees. For this binning scheme we include all the material within a vertical extent of 1.5\,kpc of the disc midplane.

Secondly, we use a vertical binning scheme where we bin the data in radius and vertical extent rather than angle. The same radial interval is used, but the $z$-direction is now binned from $-1.0$\,kpc to $1.0$\,kpc relative to the galactic midplane using 60 bins.

In each bin we record both the total H{\sc i} mass, and the H{\sc i} mass with a temperature of 200\,K or lower, which is chosen to be consistent with the CNM features identified in absorption by \citet{Heiles03}. (Although it should be noted that the majority of the gas in the CNM phase is much colder than this.) The CNM fraction $f_{C}$ is taken to be the mass of cold H{\sc i} with $T$\,$<$\,200\,K divided by the total H{\sc i} mass in the bin. The H{\sc i} surface density $\Sigma_{\rm HI}$ is then simply the total H{\sc i} mass divided by the area of the bin. 
The exact choice of temperature threshold chosen for the CNM definition makes only a small ($10-20\%$) difference to the results. For example, when testing we found the total CNM fraction for the galaxy to be $0.164$ with a threshold of 150 K, $0.192$ with 200 K, and $0.227$ for 300 K. The 200 K threshold is consequently a reasonable compromise, and has the advantage of being consistent with the observational literature.

In order to investigate the origin of the CNM fraction we calculate the mid-plane thermal pressure $P_{\rm th}$ in units of the Boltzmann constant, $k_B$, using the below sum for all cells in the bin with $|z_i|$\,$<$\,$50$\,pc in the radially binned data,

\begin{equation}
    P_{\rm th}/k_B = \sum{n_i T_i w_i}/\sum{w_i}, 
\end{equation}
where $n_i$ is the cell number density, $T_i$ the cell temperature, and $w_i$ the weighting variable, which can be either mass or volume. Note, that we investigated changing the threshold in $z$ at which material was included and found it made little difference to the average below a threshold of 100\,pc. 
Following the approach of \citet{KimCG13} we exclude dense regions with number density $n_i$\,$>$\,$50$\,cm$^{-3}$ from the average. Above such densities regions may become self-gravitating, at which point their pressure would no longer be reflective of the global disc conditions.


We also calculate the vertical turbulent pressure $P_{z}$ by taking the sum
\begin{equation}
    P_z/k_B = \frac{1}{k_B} \sum{ \rho_i vz_{i}^2 w_i} /\sum{w_i}
\end{equation}
over all cells in the midplane defined above, where $\rho_i$ is the cell density, $vz_i$ is the vertical velocity, and $w_i$ is the weighting variable of the cell. We divide $P_z$ by $k_B$ for comparison with the thermal pressure. 

To investigate the scale height of the H{\sc i} and CNM we use our vertically binned data and fit a Gaussian as was recently done by \citet{Bacchini19} for THINGS galaxies, using the equation
\begin{equation}\label{eq:scale}
    n_j = n_0 \exp {-(z_j^2/2 H_z^2)},
\end{equation}
where $n_j$ is the number density in the bin, $n_0$ the number density at the midplane, $z_j$ the distance from the midplane, and $H_z$ the scale height. The fit is performed using the \textsc{scipy} curve\_fit routine. As a second check of the vertical extent we also calculate the mass-weighted $z$  position and its dispersion, in each of our radial bins to investigate local variations.

As discussed in Section \ref{sec:sims} the simulation uses the sink particle method to track the amount of mass going into bound collapsing regions. We use these to find the growth rate of the stellar mass in each sink. To calculate this we compare the analysed time snapshot with the previous snapshot created by the simulation. If a sink has no precursor we assign it as newly created sink mass, but if it does, then we take the difference between the stellar masses to calculate how it has grown. We then divide by the time between snapshots to get the rate. The sinks are then binned spatially in the same manner as the gas mass and added to get the total star formation rate $\Sigma_{\rm{SFR}}$ from all the sinks in the bin.



\section{Results}\label{sec:results}

\subsection{Overlap with Molecular Gas}
\label{sec:molecular}

Figure \ref{fig:projection} shows a top-down view of the CNM column density alongside similar maps of the full atomic hydrogen distribution and the H$_{2}$ distribution. The CNM closely follows the distribution of the molecular gas. For the inner 12 kpc of the galaxy disc we calculate the area covering fraction of material with surface density greater than $N=10^{19}$ cm$^{-2}$ (chosen from a visual inspection of Figure \ref{fig:projection}) for the three species. H{\sc i} covered the majority of the disc ($92\%$), but the CNM phase covers only a small fraction ($5.5\%$) of the disc surface. The molecular gas covers an even smaller fraction ($3.4\%$) of the disc surface and a visual inspection of Figure \ref{fig:projection} shows that the CNM coincides more closely with the H$_2$ than the total H{\sc i}. This suggests that the CNM is both intermixed with, and extends beyond the molecular clouds. 

\begin{figure}
 \includegraphics[width = 3in]{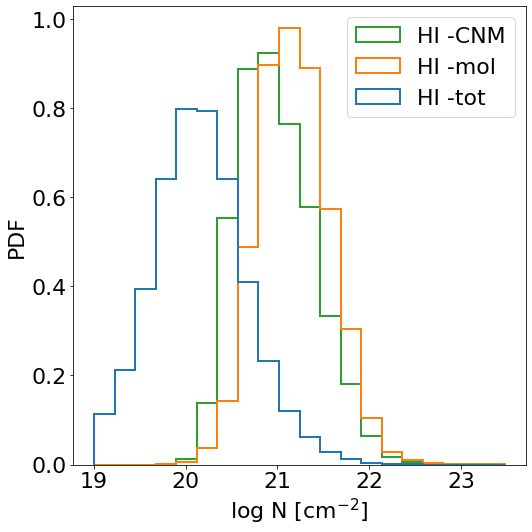}
 \caption{The probability density function distribution of H{\sc i} surface density for the \textit{total} H{\sc i}, compared to the H{\sc i} surface density where the CNM \textit{(green)} and H$_2$ \textit{(orange)} exceeded a threshold of $N=10^{19}$ cm$^{-2}$. The CNM exists in almost an identical H{\sc i} environment to the H$_2$. }
 \label{fig:HI_histogram}
\end{figure}

Figure \ref{fig:HI_histogram} shows the probability density distribution of the H{\sc i} surface density within this region. For comparison we select the pixels where the H$_2$ and CNM surface density was above our chosen threshold of $N=10^{19}$ cm$^{-2}$ and plotted the probability density of H{\sc i} surface density in these pixels. Both the CNM and molecular hydrogen are found at similar H{\sc i} surface densities. This is in good agreement with recent results from 3D dust mapping that show that nearby clouds are surrounded by extended CNM \citep{Zucker21}.

\begin{figure}
 \includegraphics[width = 3in]{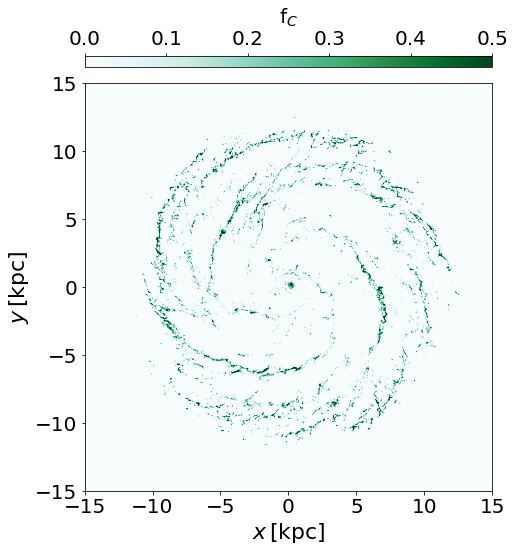}
 \caption{A map of the CNM fraction at the time corresponding to that shown in Figure \ref{fig:projection}. The CNM fraction is close to zero outside the spiral features. }
 \label{fig:fC_map}
\end{figure}

 The full H{\sc i} distribution is much more extended than the CNM and reaches out to 15\,kpc beyond the star-forming disc. The morphology of the H{\sc i} also changes between the star-forming disc and the outer regions. In the outer disc the H{\sc i} is less filamentary and fills the surface area smoothly without voids, unlike the inner regions where it follows the arms more closely. In Figure \ref{fig:fC_map} we show the CNM fraction throughout the galaxy model at the same time as shown in Figure \ref{fig:projection}. This shows that the differences between the CNM and H{\sc i} distribution are not simply due to the total mass of gas being larger in the H{\sc i} component and thus making it appear more extensive and space-filling. Instead the relative abundance of CNM to the total H{\sc i}, changes inside and outside of spiral features.

\subsection{Dependence on Galactocentric Radius}
\label{sec:radial}

\begin{figure}
 \includegraphics[width = 3in]{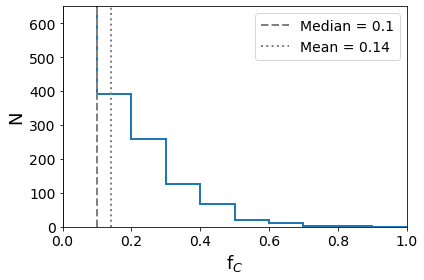}
 \caption{The distribution of the CNM fraction in all bins. The distribution is best described by a range rather than a single value.}
 \label{fig:CNM_histogram}
\end{figure}

Figure \ref{fig:CNM_histogram} shows the CNM fraction of the gas in the galaxy disc after the radial binning procedure described in Section \ref{sec:calculations}. The CNM fraction peaks at small values of $f_C$ and then gradually decreases until values of $f_C$\,$\sim$\,$0.8$. The distribution is similar to the Arecibo survey of \citet{Heiles03} but, as in that survey, a single mean value does not describe the variation in the CNM well.

\begin{figure}
 \includegraphics[width = 3.0in]{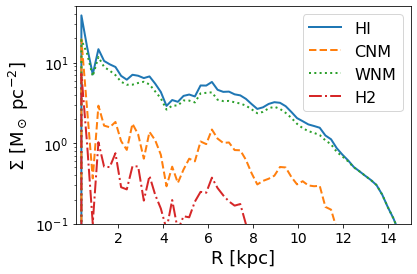}
  \includegraphics[width = 3.0in]{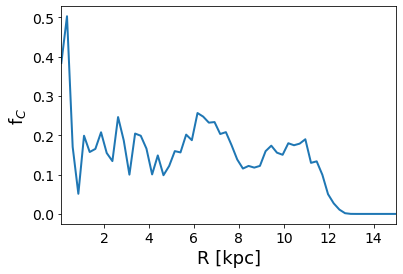}
 \caption{The radially averaged gas surface densities (\textit{top}) and CNM fraction (\textit{bottom}). The CNM fraction decreases more steeply with galactocentric radius than the total HI surface density.}
 \label{fig:CNM_radial}
\end{figure}

Figure \ref{fig:CNM_radial} shows how the H{\sc i} surface density and CNM fraction vary with galactocentric radius. As our simulated galaxy is not designed to be an exact analogue of the Milky Way we will adopt the following terminology. The `inner galaxy' is where $R$\,$<$\,$2$\,kpc, the `star-forming disc' is at $2 < R < 9$ kpc and the `outer disc' is at $R > 9$ kpc. The total H{\sc i} surface density peaks in the inner galaxy then declines radially until it is below $0.1$ M$_\odot$ pc$^{-2}$ in the outer disc. The HI surface density has multiple peaks in the star-forming disc which correspond to spiral arms. The warm and cold neutral mediums contribute to this surface density in different ways. In the inner galaxy the CNM and WNM (where here we include all H{\sc i} with $T > 200$\,K in the WNM) are well mixed and follow a similar distribution, albeit at lower surface density for the CNM. In the outer disc there is CNM only out to a radius of approximately 12 kpc in contrast to the full \HI\ distribution, which continues to beyond 14 kpc. In contrast the molecular hydrogen falls below surface densities of $0.1$ M$_\odot$ pc$^{-2}$ at a radius of 8 kpc and so is confined to the star-forming disc. Note that the H$_2$ surface density is a lower limit as some mass will be inside sink particles. \citet{Dickey09} report significant CNM in the outer disc of the Milky Way out to very large galactic radii, however we only find CNM in our model at disc radii $<12$ kpc, as opposed to $<14$ kpc for H{\sc i}. We will discuss this further in Section \ref{sec:spiral}. 

The CNM fraction is approximately constant in the disc at a value of roughly $f_C$\,$=$\,$0.2$, apart from a peak at the galactic centre. This trend of constant $f_C$ with galactocentric radius is in agreement with the absorption studies of \citet{Dickey09,Dickey22}, but is in tension with the GOTC+ survey of \citet{PinedaJ13} who see a clear radial decreasing trend in the CNM fraction derived using the C{\sc ii} line.


\begin{figure}
 \includegraphics[width = 3.4in]{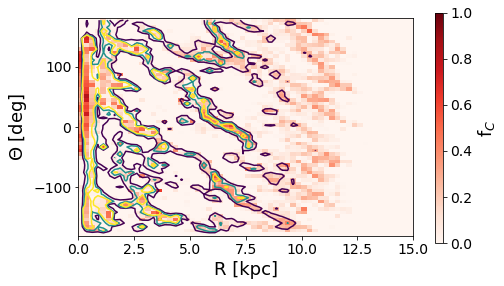}
 \caption{The CNM fraction $f_{C}$ calculated in radial and angular bins. The contours show the HI column density with contour levels of 5, and 10 M$_\odot$ pc$^{-2}$ respectively. The CNM fraction is higher in the spiral arms but is still substantial at 10 kpc where the arms are less defined.}
 \label{fig:CNM_rtheta_map}
\end{figure}

 Figure \ref{fig:CNM_rtheta_map} shows the angular dependence of the CNM fraction at increasing radii. The contours denote the H{\sc i} surface density and take the form of diagonal stripes in this phase space due to the spiral arms. Unsurprisingly, there is a low CNM fraction outside the spiral arms where the H{\sc i} column density is low and the gas temperature is hotter. However, the CNM fraction remains non-negligible between 10-12 kpc where the spiral arms are less defined.

\begin{figure*}
\begin{tabular}{c c c}
 \includegraphics[width = 2.25in]{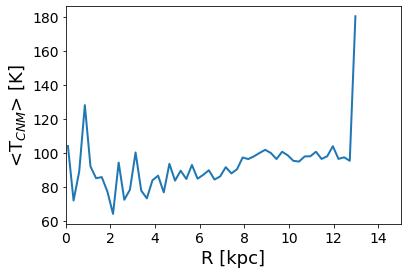}
 \includegraphics[width = 2.25in]{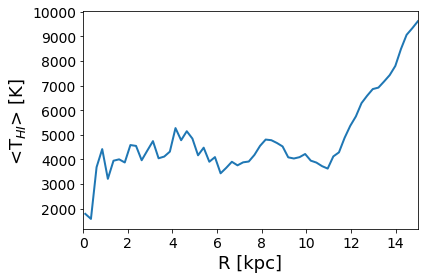}
 \includegraphics[width = 2.5in]{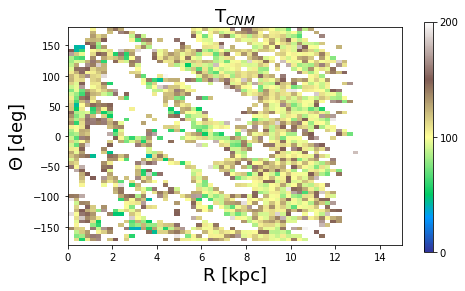}
 \end{tabular}
 \caption{The mass-weighted mean temperatures as a function of radius for the CNM (\textit{left}), and total H{\sc i} (\textit{middle}). The temperatures of both are roughly constant with galactocentric radius. The map of the mass-weighted mean CNM temperature (\textit{right}) shows that there is angular variation and that the temperature is lower in the spiral arms.}
 \label{fig:T_profiles}
\end{figure*}

When comparing to observations it is useful to understand the temperature distribution of the gas. Most obviously because the CNM, WNM split is a division in temperature, but also because the CNM temperature, $T_{\rm{cool}}$,  will determine the observed spin temperature that is directly measured from absorption features. For example, \citet{Strasser07} found that the values of $T_{\rm{cool}}$ derived from absorption features in the outer galaxy showed no strong dependence on galactic radius. In Figure \ref{fig:T_profiles} we show the mass-weighted mean CNM temperatures in our model. In agreement with observations we see a mostly flat trend. There is some hint that in the star forming disc the temperature is $\sim 10 \%$ lower than the outer disc but this is only a small variation. In the star-forming disc, the right hand panel of Figure \ref{fig:T_profiles} shows that the temperatures are cooler in the spiral arms. 

\subsection{Dependence on Vertical Extent}
\begin{figure}
 \includegraphics[width = 3.5in]{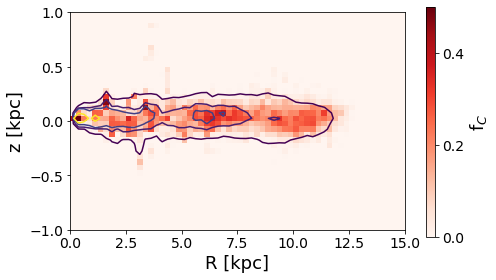}\\
  \includegraphics[width = 3.1in]{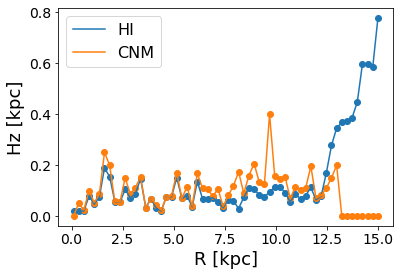}
 \caption{(\textit{Top}) The CNM fraction binned vertically as a function of galactocentric radius. The contours show the H{\sc i} column density with contour levels at 0.1, 0.5, 1.0, and 5.0 M$_\odot$ pc$^{-2}$ respectively. (\textit{Bottom}) The fitted scale height of the total H{\sc i} and the CNM as a function of galactocentric radius.}
 \label{fig:CNM_rzmap}
\end{figure}

In the top panel Figure \ref{fig:CNM_rzmap}, instead of binning with angle we bin in height above the disc. Material with a significant CNM fraction pervades all the \HI\ contours, even at the lowest surface density of $0.1$ M$_\odot$ pc$^{-2}$. To investigate this further, in the bottom panel of Figure \ref{fig:CNM_rzmap} we plot how the scale-height varies with radius in the total H{\sc i} compared to the H{\sc i} in the CNM phase. The scale height is determined by fitting to Equation \ref{eq:scale} as described in Section \ref{sec:calculations}. We find typical H{\sc i} scale heights of 100\,pc in the star forming disc, which then flares in the outer galaxy as expected from observations \citep[e.g.][]{Yim14,Bacchini19,Randriamampandry21,Soler22} The distribution is noisy due to the spiral structure but the two populations follow each other closely. However, at 12 kpc they diverge due to the CNM disappearing from the gas phase. Unlike the \HI, the CNM scale height does not flare.


\begin{figure*}
\begin{tabular}{c c}
 \includegraphics[width = 2.7in]{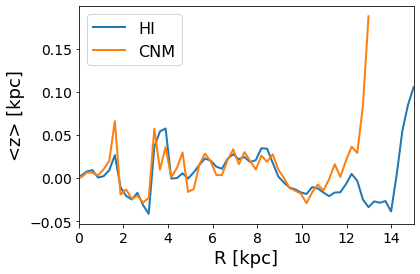}
  \includegraphics[width = 2.7in]{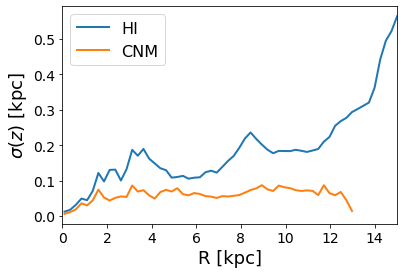}\\
   \includegraphics[width= 2.5in]{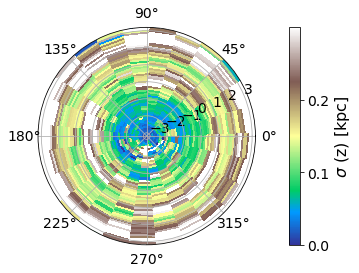}
   \includegraphics[width= 2.5in]{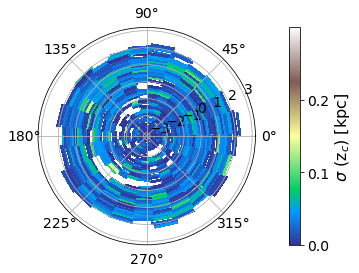}
 \end{tabular}
 \caption{(\textit{Top}): The mass-weighted mean vertical ($z$) position and its mass-weighted dispersion. (\textit{Bottom}): Maps of the dispersion in the position of the total H{\sc i} (\textit{left}), and the CNM (\textit{right}). Unlike the total H{\sc i} distribution, the CNM is tightly clustered around its mean position.}
 \label{fig:mean_z}
\end{figure*}

To investigate whether this behaviour is due to averaging we investigate another metric which does not depend on a fit. Figure \ref{fig:mean_z} shows radial profiles and angular maps of the mass-weighted mean $z$ position and its dispersion. While the mean $z$ position is always close to the 0 position in the simulation there are variations of up to 50\% of the scale height in Figure \ref{fig:CNM_rzmap}. The CNM's mean vertical position closely follows that of the full H{\sc i} distribution, but the dispersion shows very different behaviour. In the galaxy centre both the total H{\sc i} and the CNM have very low scale heights ($\sim$ 50\,pc). However, while the dispersion of the total H{\sc i} gas steadily rises, the CNM remains tightly confined below values of 100\,pc. Note that this is in good agreement with the findings of \citet{Dickey22} for the inner galaxy and star-forming disc. The CNM distribution, therefore, seems to consist of discrete `clumps' of CNM which are at a variety of vertical heights. 


\section{Discussion}
\label{sec:discussion}

\subsection{Local Pressure Environment and Star Formation}
\label{sec:pressure}

\begin{figure}
 \includegraphics[width=2.5
 in]{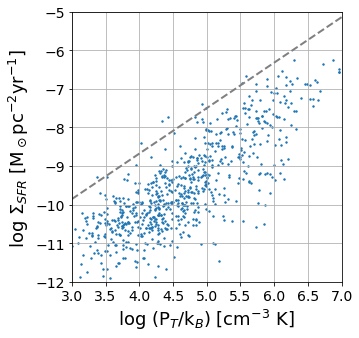}
 \caption{The total gas pressure $P_{T}$ weighted by mass vs the local star formation rate surface density $\Sigma_{\rm{SFR}}$. The grey dashed line shows the relationship found by \citet{KimCG13}.}
 \label{fig:PT_starformation}
\end{figure}

To investigate how the local pressure environment corresponds to star formation and how this connects to the CNM fraction, we plot in Figure \ref{fig:PT_starformation} the mass-averaged total (thermal and turbulent) pressure, $P_T$, vs the local star formation rate surface density $\Sigma_{\rm{SFR}}$. The grey dashed line shows the empirically determined scaling found from the simulations of \citet{KimCG13}
\begin{equation}
    \Sigma_{\rm{SFR}} =2.1\times 10^{-9} \left(\frac{P_T/k_B}{10^4}\right)^{1.18}
\end{equation}
where $\Sigma_{\rm{SFR}}$ is in units of M$_\odot$ pc$^{-2}$ yr$^{-1}$ and $P_T/k_B$ in units of cm$^{-3}$ K. This follows the same trend as the scaling of our models confirming that the total midplane pressure plays a major role in setting the global star formation rate. The normalisation of our model is different, however it should be noticed that our treatment of star formation, using sink particles, is different from the star particle approach of \citet{KimCG13} who assume a constant efficiency per free fall time and this may account for the discrepancy. However, the observational results of \citet{Sun20} are in good agreement with \citet{KimCG13} so further investigation may be needed. We also considered the volume-weighted pressure relation and found it had similar scaling behaviour but there was a discontinuous jump in $\Sigma_{\rm{SFR}}$ at high stellar densities. This is most likely due to the small volume fraction of dense star-forming regions. For this reason we consider only mass-weighted pressures going forward. Regardless of how calculated, the dispersion in the relationship decreases at large star formation rate surface densities as high pressures are needed to create sufficient dense gas. 

\begin{figure*}
\begin{tabular}{c c c}
 \includegraphics[width = 2.25in]{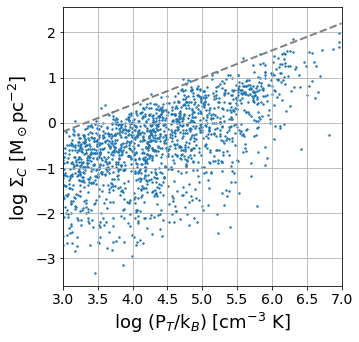}
  \includegraphics[width = 2.25in]{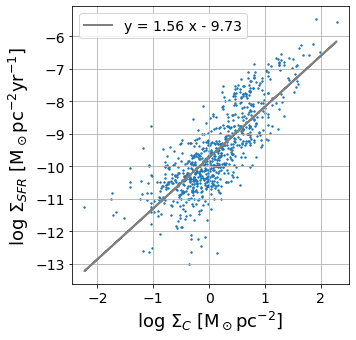}
  \includegraphics[width = 2.25in]{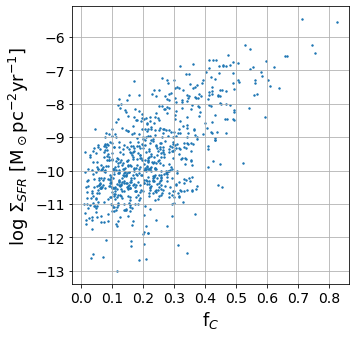}
 \end{tabular}
 \caption{\textit{Left}: The correspondence between the local mass-weighted pressure and the CNM surface density $\Sigma_C$. \textit{Middle}: The surface density of the CNM vs the local star formation rate surface density, analogous to a Kennicutt-Schmidt relation for the CNM. The grey line shows a linear fit to the data, which has a similar exponent to that derived observationally for total column density. \textit{Right}: The CNM fraction vs star formation rate surface density, which shows that higher CNM fractions correspond to greater star formation rate surface density.}
 \label{fig:Pressure_CNM}
\end{figure*}

In Figure \ref{fig:Pressure_CNM} we investigate how the total mass-weighted pressure corresponds to the CNM and star formation. The left panel shows the mass-weighted total pressure vs the CNM surface density. The dashed line shows a line with gradient of $0.6$, which matches the upper envelope of the data. There is a large amount of scatter, but the CNM column density increases with pressure as would be expected.

The middle panel of Figure \ref{fig:Pressure_CNM} shows the star formation surface density plotted against the CNM surface density in a manner analogous to a Kennicutt-Schmidt relation \citep{Kennicutt98}. Observationally, the star formation density scales with a power of $\sim1.4$ against the total gas surface density (\HI\,$+$\,H$_2$), as shown, for example, in \citet{delosReyes19}. However, when compared to only molecular gas, the scaling is linear down to at least $0.1$ solar metallicity \citep{Bigiel08,WhitworthD22}. Using the \textsc{SciPy} linear regression feature we fit a power law to the data in logspace and found a slope of 1.56 with a standard error of 0.06 best described the data. That this value is closer to the Kennicutt total gas exponent scaling with star formation than the molecular value, suggests that the CNM is not gravitationally bound and actively forming stars.

The right-hand panel of Figure \ref{fig:Pressure_CNM} shows how the CNM fraction $f_C$ corresponds to the local star formation rate surface density. There is a large scatter, particularly at low CNM fractions, however, the trend is clear that higher values of $f_C$ correspond to higher star formation rate surface densities. In the regions of our galaxy model with the most star formation, the CNM fraction is as high as $80\%$.


\begin{figure}
 \includegraphics[width = 2.5in]{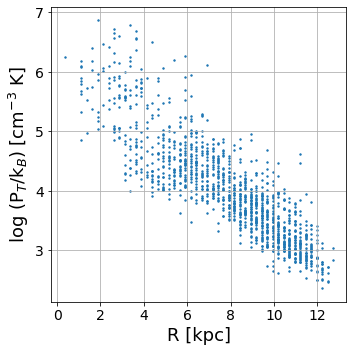}
 \caption{The galactocentric radius and mass weighted pressure of the points shown in Figure \ref{fig:Pressure_CNM} for all points where $f_C$\,$>$\,$0$. The CNM distribution ends at $R\sim 12$\,kpc, which is where the pressure falls below values of $10^{3}$\,K\,cm$^{-3}$.}
 \label{fig:radial_pressure}
\end{figure}

We can now investigate how the pressure contributes to the large-scale trends discussed in Section \ref{sec:radial}. Figure \ref{fig:radial_pressure} shows the mass-weighted pressure as a function of the galactic radius where the CNM fraction $f_C$\,$>$\,$0$. As expected the pressure steadily falls from the inner galaxy to the outer, albeit with local variations. However, strikingly the CNM vanishes from the gas phase at radii of about 12\,kpc, which corresponds to where the local pressure falls below 1000\,K\,cm$^{-3}$.

\subsection{Dependence on Radiation Field}
\label{sec:field}

\begin{figure}
 \includegraphics[width = 2.7in]{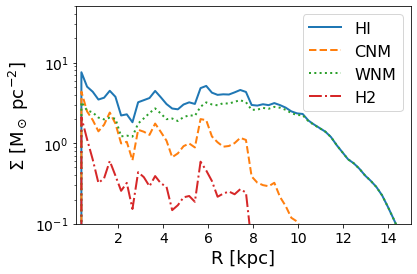}\\
  \includegraphics[width = 2.7in]{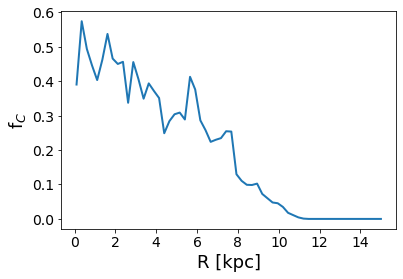}
  \includegraphics[width = 2.7in]{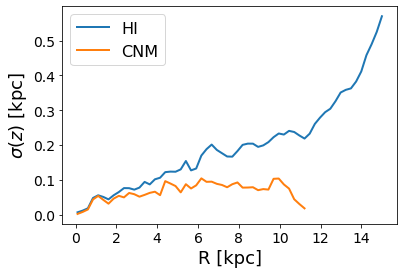}
 \caption{The radial distribution of the CNM in a galaxy model with a constant interstellar radiation field. The CNM no longer extends into the outer galaxy and the CNM fraction now decreases with galactocentric radius.}
 \label{fig:UVfield_comp}
\end{figure}

The original Isolated galaxy modelled in \citet{Tress20a} had a constant UV field and so it is possible to use this as a comparison to investigate the importance of the interstellar radiation field in setting the CNM distribution of the galaxy. Figure \ref{fig:UVfield_comp} shows the radial dependence of the CNM surface density, fraction, and vertical dispersion for the constant field of magnitude $G_0$. Most strikingly the CNM only extends to a radius of 10 kpc due to the higher field at large radii. This is the same extent as the star-forming disc in these models, and so the CNM no longer extends into the outer galaxy. The surface density of the HI and CNM are flatter as a function of radius than the surface distribution with the varying field shown in Figure \ref{fig:CNM_radial}. Intriguingly, while the radial extent of the CNM changes (from 12 to 10 kpc) the \HI\ and H$_2$ extent remains largely unchanged (at 14, and 8 kpc respectively). Clearly, the CNM is far more sensitive to the local radiation field in this regard compared to the other phases.

With a constant interstellar radiation field the CNM fraction $f_C$, shown in the middle panel of Figure \ref{fig:UVfield_comp}, decreases radially rather than remaining flat in the disc as was the case in Figure \ref{fig:CNM_radial}. The vertical dispersion of the CNM, however, remains similar with and without the varying field with the CNM having a dispersion of around 100 pc throughout the disc in both cases with no flaring.

\begin{figure}
  \includegraphics[width = 2.25in]{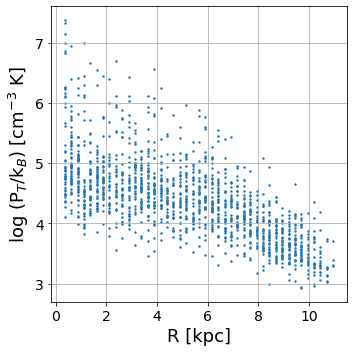}
  \caption{The galactocentric radius and mass-weighted pressure for a model with constant interstellar radiation field for all bins with $f_C>0$. The CNM distribution now ends at $R\sim 10$\,kpc, which is where the pressure falls below values of $10^{3}$\,K\,cm$^{-3}$ in this model.}
 \label{fig:UVfield_pressure}
\end{figure}

To explore the origin of this behaviour we plot the mean total midplane pressure as a function of radius for the model with a constant field in Figure \ref{fig:UVfield_pressure}. The pressure now falls below a value of $10^{3}$\,K\,cm$^{-3}$ at a galactocentric radius of 10 kpc as opposed to 12 kpc in the varying case. Both of these radii exactly correspond to where the CNM disappears from the gas phase, once again confirming the important role of midplane pressure in setting the CNM distribution.

\subsection{Turbulence in the Outer Galaxy}
\label{sec:turb}

However, pressure alone is not a full explanation for the distribution of the CNM in the gas phase. Turbulent velocity fluctuations are important for pushing gas out of equilibrium and creating local over-densities where the CNM can form. The vertical velocity dispersion observed in \HI\ in spiral galaxies is observed to be both highly turbulent, and varying radially from values of roughly $12$ to $15$\,kms$^{-1}$ in the central parts, to $\sim 4-6$ kms$^{-1}$ in the outer parts e.g. \citep{Dickey90,Kamphius93,Rownd94,Meurer96,deBlok06}. Such velocity dispersions can be theoretically explained via a number of mechanisms that include supernova feedback \citep{Dib06}, MRI instability \citep{Piontek05,Piontek07}, or accretion from the galactic halo \citep{Klessen10}. However, none of these effects are operating in our model. In the outer galaxy there is no star formation and hence no supernova feedback. Our model is purely hydrodynamic, and it is isolated with no accretion of material from the intergalactic medium.

\begin{figure}
  \includegraphics[width = 2.7in]{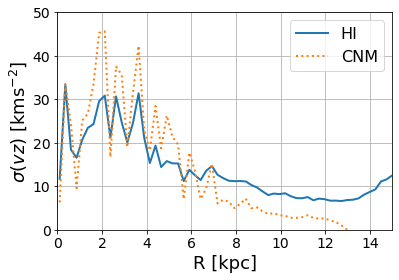}
  \caption{The mass-weighted radial average of the vertical velocity dispersion. The dispersion follows the observed trend of higher values in the inner galaxy and star-forming disc, but still substantial vertical dispersion in the outer galaxy.}
 \label{fig:sigmav}
\end{figure}

In Figure \ref{fig:sigmav} we plot the mass averaged vertical velocity dispersion in radial bins for \HI\ and the CNM. For both phases the velocity dispersion is higher in the star-forming disc where we have supernova feedback, however, it remains substantial and of similar magnitude to observations in the outer galaxy. Therefore we must conclude that galactic dynamics alone are enough to match the observed velocity dispersion. Consequently, we are confident that we are not underestimating the CNM fraction in our models due to a lack of turbulence.

\subsection{CNM and Spiral Arms}
\label{sec:spiral}

In our model, the CNM extends to midway in the outer disc but does not cover its full extent. However, the absorption studies of \citet{Dickey22} find CNM in the entirety of the outer disc of the Milky Way, out to radii of $40$ kpc. Kinematic distances are hard to estimate and absorption measurements give information along single sight lines. Still, it seems clear that in the Milky Way the CNM can exist beyond the radii predicted in our models.

\begin{figure}
 \includegraphics[width = 2.8in]{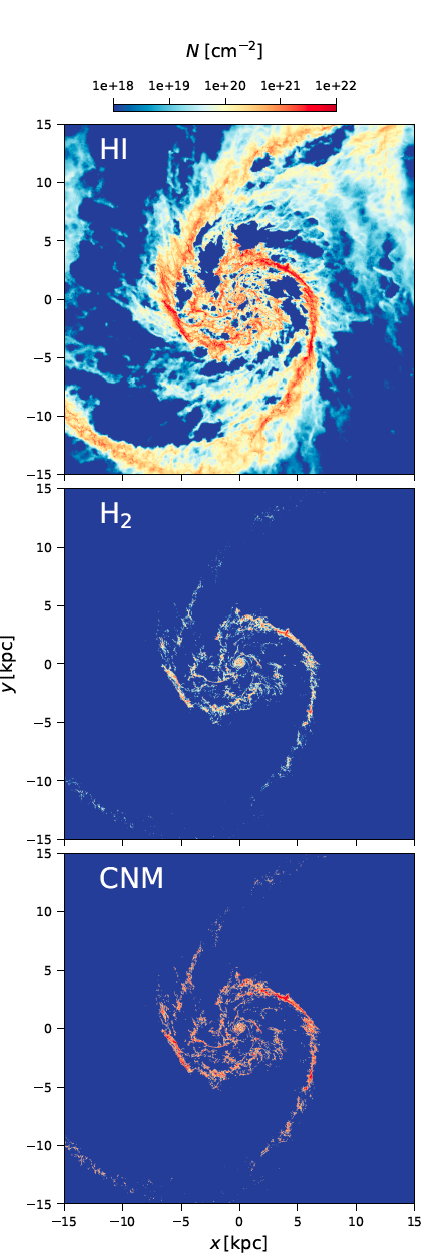}
 \caption{Projected column density of the interacting simulation in H{\sc i}, H$_2$, and the CNM. The CNM now extends into the outer galaxy.}
 \label{fig:proj_spiral}
\end{figure}

One possibility is that spiral arm structure is able to concentrate gas in isolated pockets in the outer disc. For example, \citet{Strasser07} observe CNM in the outer Milky Way, but specifically target spiral arms. Our fiducial model is of an isolated galaxy without strong spiral structure in the outer regions. Therefore to test the importance of spiral arms in forming the CNM we turn to the other model described by \citet{Tress20a}, where a close encounter has triggered the formation of well-defined spiral arms to create an `M51 analogue'. Figure \ref{fig:proj_spiral} shows the gas distribution of the Interacting model at the same time as the isolated model analysis. Note that the interstellar radiation field is constant in this model so we also compare it to the model discussed in Section \ref{sec:field} rather than just our fiducial case. Figure \ref{fig:proj_field} in Appendix \ref{sec:fieldmaps} shows the column density maps for the Isolated model with a constant interstellar radiation field.

The atomic hydrogen distribution in the outer galaxy is now very different from the Isolated constant field case. Where before the distribution smoothly filled the surface of the outer disc, now the atomic hydrogen is concentrated in the extended spiral arms. The lower panels of Figure \ref{fig:proj_spiral} show that both H$_2$ and the CNM now can be found in the outer galaxy. For instance, a particularly large clump of CNM can be seen in the lower left corner. Once again the CNM closely follows the molecular hydrogen, but with a higher column density over a slightly larger area than the Isolated case.

\begin{figure}
 \includegraphics[width = 3.2in]{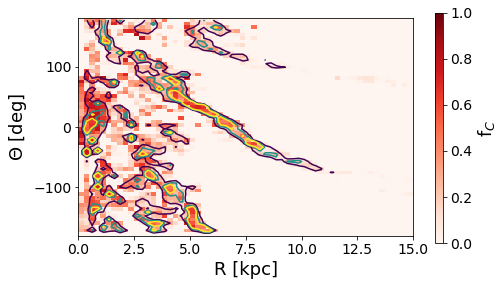}
 \caption{The CNM fraction $f_{C}$ calculated in radial and angular bins for the spiral arm simulation. The contours show the HI column density with contour levels of 5, 10, and 15\,M$_\odot$\,pc$^{-2}$ respectively. Again the CNM is concentrated in the spiral arms.}
 \label{fig:interact_CNM_maps}
\end{figure}

\begin{figure}
 \includegraphics[width = 3.0in]{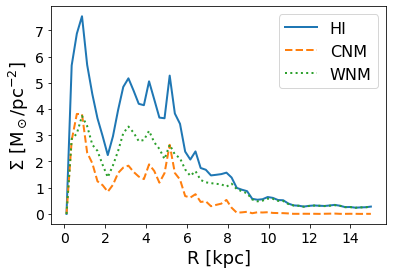}
  \includegraphics[width = 3.0in]{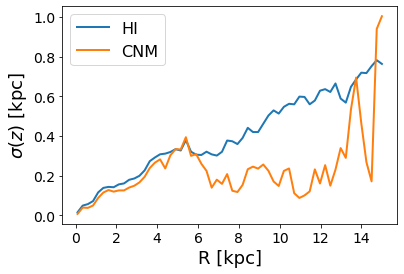}
 \caption{\textit{Top} The radially averaged H{\sc i} surface density for the spiral-arm mode. \textit{Bottom} the mass-weighted dispersion about the vertical ($z$) position as a function of radius. }
 \label{fig:interact_radial}
\end{figure}

Figure \ref{fig:interact_CNM_maps} shows the binned CNM fraction of the interacting galaxy. In the spiral arms, the CNM now persists to large radii albeit at a low level. Figure \ref{fig:interact_radial} shows the radially averaged profiles for these maps. When averaged over an annulus of the galaxy disc the surface density of the CNM is negligible in the outer galaxy and follows the same radially decreasing trend as Figure \ref{fig:CNM_radial}. 

Finally, the lower panel of Figure \ref{fig:interact_radial} shows the mean dispersion in the z position, $\sigma(z)$, of both H{\sc i} and the CNM. In the inner galaxy, the CNM and H{\sc i} have increasing dispersion, suggesting that they are both well mixed vertically in the disc. However, beyond 6\,kpc, where the dense gas is mainly in spiral arms, $\sigma(z)$ drops for the CNM but continues to rise for H{\sc i}. This further suggests that the CNM in the extreme outer galaxy is confined to distinct clumps of gas where the local pressure has been enhanced by spiral arms.

Certainly, the scenario presented here is only one possibility for how CNM might exist in the extreme reaches of the outer galaxy. This paper focuses on a pair of models, and is not a parameter study. It explores two extremes, one model with weak spiral structure, and another with strong, and consequently brackets a range of possible CNM distributions. Magnetic fields could add additional support to molecular cloud envelopes which could then contain a substantial amount of CNM as well as CO-dark molecular gas \citep[e.g][]{Smith14a}. Another possibility is that cold gas could be accreted onto the outer galaxy from the circumgalactic medium such as in the Magellanic Stream of our own Galaxy \citep{Fox14}. Alternatively, the discrepancies that we see between Milky Way observations and our model may be due to projection effects. Our analysis has been from an external perspective, in order to better quantify the variation with galactocentric radius and vertical height above the disc midplane within the Milky Way the CNM distribution has to be recovered from sight lines containing emission and absorption from gas at multiple different radial positions within the disc.

\section{Conclusions}
\label{sec:conclusions}

We have investigated the distribution of the CNM in new high-resolution simulations of an isolated disc galaxy based on those presented by \citet{Tress20a} but with a radially varying interstellar radiation field. Our \textsc{Arepo} simulations use a variety of custom modules allowing us to follow the chemical and star-forming evolution of the dense gas. These included time-dependent gas chemistry, gas self-shielding from the ambient UV field, sink particles to represent star formation, and supernova feedback. The resulting models allow the distribution of the CNM to be followed down to sub-pc scales in the dense gas. Most of our analysis focuses on an isolated galaxy model, however, we later compare it to a model with a constant interstellar radiation field, and another with identical mass which has been perturbed by a close encounter to generate strong spiral arms that extend to the outer galaxy.

Our conclusions are as follows:

\begin{enumerate}
\item No single value describes the CNM fraction $f_C$ (the fraction of the atomic gas with T<200 K relative to the total \HI) everywhere in the galaxy. Values range from $f_C=0$ to $0.8$ with high values being less likely.

\item The CNM is not uniformly distributed in the star-forming disc, but follows a clumpy distribution. A comparison of the column density maps shows that it overlaps more closely with the H$_2$ distribution than the total \HI. This is particularly true in spiral arms where the CNM fraction is clearly enhanced.

\item The CNM extends into the outer galaxy, beyond the star-forming disc but does not extend as far out as the full \HI\  distribution. In our model, the radial surface density profiles of \HI, the CNM, and H$_2$ remain above a value of $0.1$ M$_{\odot}$ pc$^{-2}$ out to radii of 8, 12, and 14 kpc respectively.

\item The CNM fraction remains approximately constant with galactocentric radii in agreement with measurements with \HI\ absorption. This effect is a consequence of the falling interstellar radiation field compensating for the decreasing column density at large radii. In our comparison model with a constant interstellar field the CNM fraction decreased with galactocentric radius.

\item The vertical distribution of the CNM is clumpy in the star-forming disc with a scale height of around $\sim 100$ pc and no flaring. An analysis of the dispersion, $\sigma(z)$, suggests that the CNM is more localised in the $z$-direction than the overall H{\sc i} distribution.

\item The star formation rate in the galaxy is well correlated with the total midplane pressure in logspace as suggested by \citet{KimCG13}. We find that the CNM column density also correlates well with pressure as expected from models of the ISM such as \citet{Wolfire03}. We find no CNM in our isolated galaxy discs beyond the point where the total pressure $P_T/k_B$ drops below $1000$\,K\,cm$^{-3}$.

\item The `CNM Kennicutt-Schmidt' relation (i.e. the star formation rate surface density plotted against the CNM column density) has a scaling of $\sim 1.5$, which is more similar to the relation seen for total column density than the linear relation seen for molecular gas. This suggests that while the formation of the CNM is a precursor to star formation, it is not the predictor of gravitationally collapsing star-forming regions that the H$_2$ is. However, the CNM fraction does increase with star formation density, with the highest CNM fractions always associated with active star formation. 

\item The vertical velocity dispersion of the \HI\  in our models is in good agreement with observations of \HI\  in nearby galaxies. In the star-forming disc a major contribution to this is supernova feedback, but even in the outer galaxy where there is no star formation galactic dynamics is sufficient to drive velocity dispersions between $5-10$ km s$^{-1}$.

\item Spiral arm features in the outer galaxy may give rise to isolated clumps of CNM at extremely large radii, beyond where it is expected in our more symmetric isolated models. This may be an explanation for the CNM seen at extremely large galactic radii in the Milky Way by \HI\  absorption studies \citep{Dickey22}.

\end{enumerate}

\section*{Acknowledgements}

RJS gratefully acknowledges an STFC Ernest Rutherford fellowship (grant ST/N00485X/1).
RSK, SCOG, JDS, SM, and DW acknowledge support from the DFG via the collaborative research center (SFB 881, Project-ID 138713538) “The Milky Way System” (subprojects A1, B1, B2 and B8), from the Heidelberg Cluster of Excellence “STRUCTURES” in the framework of Germany’s Excellence Strategy (grant EXC-2181/1, Project-ID 390900948) and from the European Research Council (ERC) via the ERC Synergy Grant “ECOGAL” (grant 855130). RSK furthermore thanks the German Ministry for Economic Affairs and Climate Action for funding in the project ``MAINN'' (funding ID 50OO2206).  M-MML acknowledges partial funding from NSF grant AST18-15461.
%
%
This work used the DiRAC@Durham facility managed by the Institute for Computational Cosmology on behalf of the STFC DiRAC HPC Facility (www.dirac.ac.uk). The equipment was funded by BEIS capital funding via STFC capital grants ST/P002293/1, ST/R002371/1 and ST/S002502/1, Durham University and STFC operations grant ST/R000832/1. DiRAC is part of the National e-Infrastructure.
The Heidelberg group also acknowledges HPC resources and data storage supported by the State of Baden-W\"{u}rttemberg (MWK) and DFG through grant INST 35/1314-1 FUGG and INST 35/1503-1 FUGG, and for computing time from the Leibniz Computing Centre (LRZ) in project pr74nu.

\section*{Data Availability}

Both the simulation data, projection maps and help with reading in the data is available on request by contacting the first author.



\bibliographystyle{mnras}
\bibliography{bibliography.bib} 




\appendix

\section{CNM maps with a constant radiation field}
\label{sec:fieldmaps}

\begin{figure}
 \includegraphics[width = 2.8in]{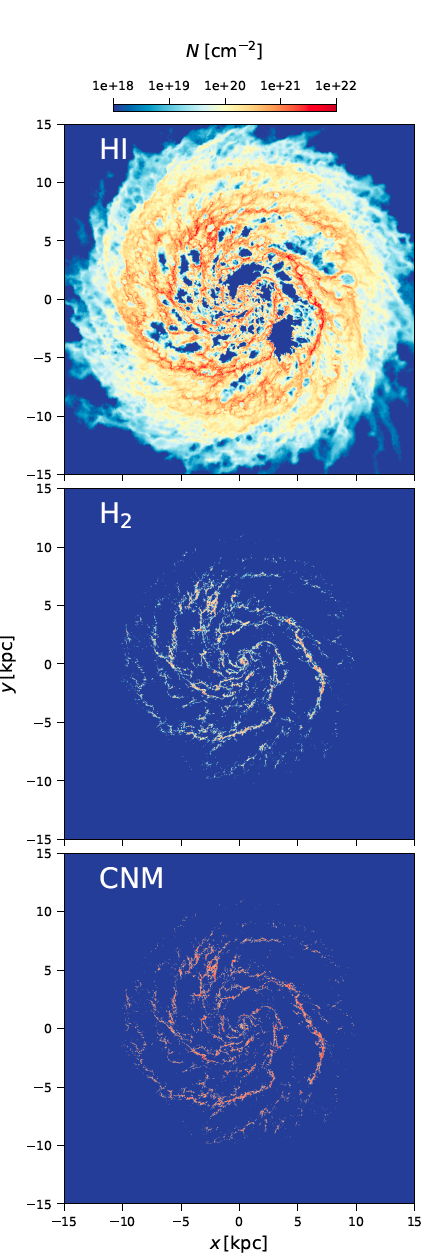}
 \caption{Projected column density of H{\sc i}, H$_2$, and the CNM in the Isolated models with a constant interstellar radiation field. The CNM traces the distribution of the molecular gas even more tightly and no longer extends to the outer disc.}
 \label{fig:proj_field}
\end{figure}

In Figure \ref{fig:proj_field} we show the projected column density maps for H{\sc i}, H$_2$, and the CNM in the Isolated models with a constant interstellar radiation field. The CNM no longer extends to the outer galaxy but is tightly confined to around the H$_2$ in the star-forming disc.





\bsp	
\label{lastpage}
\end{document}